\documentstyle[aps,prb]{revtex}

\begin{document}
\draft
\title{ Modulation structure of a frustrated spin ladder}
\author{Yupeng Wang}
\address{
Institute of Physics \& Center for Condensed Matter Physics, 
Chinese Academy of Sciences, 
Beijing 100080, People's Republic of China}
\maketitle
\begin{abstract}
We study a two-leg spin-1/2 ladder with isotropic exchanges and biquadratic interactions
in the basic plaquettes. It is shown that for the extremely frustrated case, 
the system exhibits a self-organized 
phase separation. In some parameter regions, the singlet rungs form a Wigner-like lattice in
the triplet-rung host. There are three types of 
elementary excitations in this modulation phase, i.e., the spinons in a triplet domain, the broken
singlet rungs and the deformation of the Wigner-like lattice. The flux phase induced 
by an external magnetic field in the rung-dimerized phase is also discussed.
\end{abstract}
\vspace{0.5cm}
After the discovery of the ladder superconductors, the current interest in the coupled
spin chains has been greatly renewed.\cite{1} It is well established that the regular
spin-1/2 ladders with even number of legs have a gaped  spin-liquid ground state, while
the odd-legged ladders have a gapless spin-liquid ground state. On the other hand, the generalized
spin ladders including other couplings beyond the nearest neighbor exchanges, which can
interpolate among a variety of systems, have been demonstrated exhibiting remarkably rich
behavior.\cite{2,3,4,5} The hamiltonian of the generalized spin ladder reads:
\begin{eqnarray}
H=\frac14J_1\sum_{j=1}^N(\vec{\sigma}_j\cdot\vec{\sigma}_{j+1}+\vec{\tau}_j\cdot\vec{\tau}_{j+1})
+\frac14J_2\sum_{j=1}^N(\vec{\sigma}_j\cdot\vec{\tau}_{j+1}+\vec{\tau}_j\cdot\vec{\sigma}_{j+1})
\nonumber\\
+\frac12J_3\sum_{j=1}^N\vec{\sigma}_j\cdot\vec{\tau}_j
+\frac14U_1\sum_{j=1}^N(\vec{\sigma}_j\cdot\vec{\sigma}_{j+1})
(\vec{\tau}_j\cdot\vec{\tau}_{j+1})\\+\frac14U_2\sum_{j=1}^N(\vec{\sigma}_j\cdot\vec{\tau}_{j+1})
(\vec{\sigma}_{j+1}\cdot\vec{\tau}_{j})+\frac14U_3\sum_{j=1}^N(\vec{\sigma}_j\cdot\vec{\tau}_j)
(\vec{\sigma}_{j+1}\cdot\vec{\tau}_{j+1}),\nonumber
\end{eqnarray}
where ${\vec\sigma}_j$ (${\vec\tau}_j$) are Pauli matrices on site $j$ of the upper (lower) leg,
$J_\alpha$ ($\alpha=1,2,3$) are the exchange coupling constants and $U_\alpha$ are the biquadratic
coupling constants. The hamiltonian (1) contains all possible $SU(2)$-invariant interactions in
a basic plaquette formed by two nearest rungs. We note the biquadratic interactions may be effectively
mediated by phonons\cite{2} and their importance for some properties of $CuO_2$ plaquette has been 
pointed out.\cite{6} In addition, current experiments have revealed that such multi-spin interactions
are realized in the two-dimensional (2D) solid $^3He$,\cite{7} 2D Wigner solid of electrons formed in a 
$Si$ inversion layer\cite{8}, and the $bcc$ solid $^3He$.\cite{9} \par
The model (1) has been studied by many authors in some parameter regions. 
The possible phases including the dimerized phase,\cite{2} the Haldane phase\cite{3,5}
and the rung-dimerized phase have been reported. In addition, the model (1) has a variety of integrable
cases \cite{10,11} and some gapless phases have been found. All these results indicate that the
generalized spin ladder has a very rich ground state phase diagram. 
In this letter, we study the extremely 
frustrated, i.e., $J_1=J_2$, $U_1=U_2$ case. We show that there are three possible phases in this case,
i.e., a rung-dimerized (RD) phase, a triplet-rung (TR) phase  interpolating between the gapless spin liquid and the
Haldane spin liquid\cite{12} or the $VBS$ state,\cite{13} and a new gapful phase with modulation
structure. The latter phase is a mixed state (MS) consisting of both singlet rungs and triplet 
rungs and its structure is very similar
to a Wigner lattice (crystallization of the singlet rungs) or the flux phase of the  type II
superconductors (with the singlet rungs as the effective fluxes). \par
In the extremely frustrated case ($J_1=J_2$, $U_1=U_2$), the hamiltonian (1) can be rewritten as
\begin{eqnarray}
H={\tilde J}\sum_{j=1}^N[{\vec S}_j\cdot{\vec S}_{j+1}-\beta({\vec S}_j\cdot{\vec S}_{j+1})^2]
\nonumber\\+U\sum_{j=1}^N(2-{\vec S}_j^2)(2-{\vec S}_{j+1}^2)
-J\sum_{j=1}^N(2-{\vec S}_j^2)+C,
\end{eqnarray}
where ${\vec S}_j=({\vec \sigma}_j+{\vec\tau}_j)/2$ is the total spin of the $j$-th rung, 
${\tilde J}=2J_1+U_1$, $\beta=-2U_1/(2J_1+U_1)$, $U=U_3$, $J=J_3+U_3-2U_1$ and $C=J_3/2+U_3/4-5U_1/2$ 
is an irrelevant constant. For convenience, we put ${\tilde J}=1$ and omit the irrelevant constant
$C$ in the following text. The total spin $S_j$ takes two possible values
$0,1$, corresponding to the singlet rungs and triplet rungs, respectively. Based on Eq.(2), the phase diagram
of the system can be conjectured roughly. There are three possible types of ground state configurations.
For a large enough $J$, the singlet rungs are more stable
than the triplet rungs. The ground state is a simple product of $N$ singlet rungs, i.e., a rung-dimerized
state. For a small $J$ and large $U$, the triplet rungs are dominant over the singlet rungs and the 
ground state is exactly the same of a generalized spin-1 chain. In some intermediate $J,{~}U$ regions,
the singlet rungs and the triplet rungs may coexist in the ground state. Since there is no genuine
interactions between the singlet rungs and the triplet rungs as we can read off from Eq.(2), 
a self-organized phase separation
occurs in this case. The singlet rungs simply cut the system into disconnected domains of triplet
rungs and a positive $U$ will stablize a Wigner-like lattice of the singlet rungs in the triplet liquid.
Each triplet domain in the mixed state behaves as an open spin-1 chain. 
Such a mixed state can only be accompanied by a negative boundary energy, as the
flux phase in a type II superconductor. We note that $\beta=\pm1$ represent two integrable
points of the model (2) and when $\beta=-1/3$, a $VBS$ ground state can be constructed.\cite{13}
Very interestingly, the translational invariance is broken in the MS phase and the ground
state is no longer a spin liquid but a periodic array of spin-liquid domains. \par
To give a clear picture,
we study one of the integrable cases $\beta=-1$. In this case, each triplet domain behaves as
an $SU(3)$-invariant spin-1 chain with open boundaries and can be solved exactly via Bethe ansatz.\cite{14,15}
Let us consider a triplet domain with $M$ rungs. 
The effective hamiltonian of this domain reads
\begin{eqnarray}
H_M=\sum_{j=1}^{M-1}[{\vec S}_j\cdot{\vec S}_{j+1}+({\vec S}_j\cdot{\vec S}_{j+1})^2]
\end{eqnarray}
with $S_j=1$. The solution of Eq.(3) is exactly the same of the $SU(3)$-invariant $t-J$ model\cite{16}
with pure open boundaries.
Its spectrum is determined by the following Bethe ansatz
equations (BAE's)
\begin{eqnarray}
\left(\frac{\lambda_j-\frac i2}{\lambda_j+\frac i2}\right)^{2M}=\prod_{r=\pm1}
\left[\prod_{l\neq j}^{M_1}\frac{\lambda_j-r\lambda_l
-i}{\lambda_j-r\lambda_l+i}\prod_{\alpha=1}^{M_2}\frac{\lambda_j-r\mu_\alpha+\frac i2}
{\lambda_j-r\mu_\alpha-\frac i2}\right],\nonumber\\
\prod_{r=\pm1}\prod_{\beta\neq \alpha}^{M_2}\frac{\mu_\alpha-r\mu_\beta-i}{\mu_\alpha-r\mu_\beta+i}
=\prod_{r=\pm1}\prod_{j=1}^{M_1}\frac{\mu_\alpha-r\lambda_j-\frac i2}{\mu_\alpha-r\lambda_j+\frac i2},
\end{eqnarray}
where $\lambda_j$ and $\mu_\alpha$ are the rapidities of the spinons and $M_2< M_1< M$. 
The eigen energy of the triplet domain reads:
\begin{eqnarray}
E_M=-\sum_{j=1}^{M_1}\frac1{\lambda_j^2+1/4}+2(M-1).
\end{eqnarray}
Notice that the second term in Eq.(5) is also relevant since the number of triplet rungs is not fixed 
in the whole system. It contributes an amount of -2 to the boundary energy, which is crucial to stablize 
an MS. For $M>>1$, the ground state energy of the triplet domain can be expressed as $E_M=M\epsilon_0+
\epsilon_b(M)$, where $\epsilon_0$ is the energy density of an infinite $SU(3)$ spin-1 chain,
$\epsilon_b(M)$ is the boundary energy including the $O(M^{-1})$ finite size correction. Denote
the ground state distributions of $\lambda$ and $\mu$ for $M\to\infty$ as $\rho_1(\lambda)$
and $\rho_2(\mu)$ respectively. From the BAE's (4) we get the following integral
equations
\begin{eqnarray}
\rho_1(\lambda)+[2]\rho_1(\lambda)-[1]\rho_2(\lambda)=\frac1{2M}a_2(\lambda)+a_1(\lambda),\nonumber\\
\rho_2(\mu)+[2]\rho_2(\mu)-[1]\rho_1(\mu)=\frac1{2M}a_2(\mu),
\end{eqnarray}
where $a_n(\lambda)=n/2\pi[\lambda^2+(n/2)^2]$ and $[n]$ is the integrable operator with kernel
 $a_n(\lambda)$.
By solving the above integral equations with
Fourier transformation,\cite{17,18,19} we obtain
\begin{eqnarray}
\rho_1(\lambda)=\rho_1^0(\lambda)+\frac1M\delta\rho_1(\lambda),\nonumber\\
\rho_1^0(\lambda)=\frac1{2\pi}\int e^{-i\omega\lambda}\frac{2\cosh\frac\omega2}{4\cosh^2\frac\omega2-1}
d\omega,\\
\delta\rho_1(\lambda)=\frac1{4\pi}\int e^{-i\omega\lambda}\frac{e^{-\frac{|\omega|}2}
}{2\cosh\frac\omega2-1}d\omega-\frac12\delta(\lambda),\nonumber
\end{eqnarray}
where $\rho_1^0(\lambda)$ is the density of $\lambda$ of the ground state for $M\to\infty$
with periodic boundary conditions. The $\delta(\lambda)$ term in the third equation of Eq.(7) comes
from the $\lambda=0$ mode, which is forbidden in an open boundary system.\cite{16,18} $\epsilon_0$
and $\epsilon_b(M)$ can be derived as
\begin{eqnarray}
\epsilon_0=-\int\frac{\rho_1^0(\lambda)}{\lambda^2+1/4}d\lambda+2=2-\ln3-\frac\pi{3\sqrt3},\nonumber\\
\epsilon_b(M)=-\int\frac{\delta\rho_1(\lambda)}{\lambda^2+1/4}-2+O(M^{-1})=\frac49\sqrt3\pi
-4+O(M^{-1}).
\end{eqnarray}
Based on Eq.(8), the phase boundaries can be determined exactly. As we discussed above, there are three
possible phases, i.e., the RD phase, the TR liquid and the MS phase
containing both singlet rungs and triplet rungs. We note that the density of the
ground state energy of the rung-dimerized state reads $\epsilon_r=4U-2J$, as we can
easily derive from Eq.(2).  Therefore, the phase
boundary between  the RD phase and the TR liquid phase is given by
$4U-2J=\epsilon_0$. On the other hand, generating a singlet rung in the TR liquid implys a broken triplet
rung and an open boundary. The excitation energy of this process is $\epsilon_b(\infty)-\epsilon_0-2J$. 
Therefore, the phase boundary between the TR phase and the mixed state  is given by 
$2J=\epsilon_b(\infty)-
\epsilon_0$. We note the singlet rungs are very similar to the fluxes in a  type II superconductor.
Here $\epsilon_b(\infty)<0$ corresponds to the surface energy, while $-2J-\epsilon_0$ corresponds to the
self energy of the fluxes. When $2J>\epsilon_b(\infty)-\epsilon_0$, some singlet rungs appear in the triplet
liquid. Here $J_{c1}=(\epsilon_b(\infty)-\epsilon_0)/2$ serves as the lower critical field.
Due to the finite size correction of $\epsilon_b(M)$, the singlet rungs are unfavorable to
close each other (note $\epsilon_b(M)$ is an increasing function\cite{17,18,19} of $1/M$). The lengths of the triplet
domains are mainly controlled by the finite size correction. Suppose we have
$N_M=N/(M+1)$ triplet domains with lengths $M+\delta_m$ respectively. The small shifts $\delta_m$
satisfy the condition $\sum_m\delta_m=0$. The total correction energy reads
\begin{eqnarray}
E_c=\sum_{m=1}^{N_M}[\epsilon_b(M+\delta_m)-\epsilon_b(\infty)]\sim\sum_{m=1}^{N_M} \frac1{M+\delta_m}.
\end{eqnarray}
By minimizing Eq.(9), we get $\delta_m=0$, which indicates a periodic
array of the singlet rungs in the whole system. This novel modulation
structure is very similar to  a Wigner lattice or an Abrikosov lattice but with a very different physical 
interpretation. As in the type II superconductors, there is also an upper bound of $J$ 
corresponding to the upper critical field, which determines the phase boundary between the 
RD phase and the MS phase. Suppose an MS is stable at $M=M(J)$. The phase boundary is given by
\begin{eqnarray}
4U-2J=\epsilon_{M(J)}+\frac1{M(J)+1}[\epsilon_{M(J)}-2J].
\end{eqnarray}
When $M(J)\to 1$, $J_{c2}\to 4U$. \par
Now we turn to the elementary excitations in the MS phase. There are three types of elementary 
excitations in this case, i.e., the spinons in the triplet domains, broken singlet rungs and the
deformation of the Wigner-like lattice. Suppose the lattice is stablized with a period $M+1$ ($M$
triplet rungs and one singlet rung). The spinons in the present case have
a finite energy gap in the order of $M^{-1}$ due to the finite length of the domains. For $M>>1$, the
energy gap can be derived with the well known finite-size-correction techniques.\cite{17,18,19} As in the
usual integrable models, the excitation energies of the spinons are given by the so-called dressed 
energies\cite{19} which in our case read (for $M\to\infty$)
\begin{eqnarray}
\epsilon_1(\lambda)=-2\pi a_1(\lambda)+[1]\epsilon_2(\lambda)-
[2]\epsilon_1(\lambda),\nonumber\\
\epsilon_2(\mu)=[1]\epsilon_1(\mu)-[2]
\epsilon_2(\mu).
\end{eqnarray}
The velocities of the two branches of low-energy spinons are equivalent, which read
\begin{eqnarray}
v=\lim_{\lambda\to\infty}\frac{\epsilon_1'(\lambda)}{2\pi\rho_1^0(\lambda)}
=\lim_{\mu\to\infty}\frac{\epsilon_2'(\mu)}{2\pi\rho_2^0(\mu)}
\end{eqnarray}
The energy gap associated with the spinon excitations is thus $\Delta\approx \pi v/M$. There are three
different pictures of the spinon excitations: (i)For $M=3n$, there is no holes in the $\lambda$- and
$\mu$-seas in the ground state. The spinons are generated by  spin flips. The simplest excitation is
a one $\lambda$-hole and two $\mu$-hole state. (ii)For $M\neq 3n$, there are some $\lambda$- and $\mu$-
holes in the ground state. The mobility of the holes gives the simplest excitation. (iii)For arbitrary
$M$, there are string excitations. We note all these excitations have the same energy gap.\par
A broken singlet rung in the modulation phase indicate that two neighboring triplet domains combined
to a single domain with length $2M+1$. In a stable MS phase, $M$ is determined by minimizing
the per site energy $\epsilon_M=[\epsilon_b(M)-\epsilon_0-2J]/(M+1)+\epsilon_0$. For large $M$,
from $\partial\epsilon_M/\partial M=0$ we readily obtain $J-[\epsilon_b(\infty)-\epsilon_0]/2\sim 
M^{-1}$. Therefore, the excitation energy of a broken singlet rung is $\epsilon_b(2M+1)-\epsilon_b(M)
+\epsilon_0+2J$, which is still in the order of $M^{-1}$. \par
The deformations of the Wigner-like lattice represent another type of excitations in the modulation
phase. This type of excitations are static rather than dynamic due to the complete phase separation.
Suppose one singlet rung is moved from its equilibrium position by $\delta M$. The two neighboring
triplet domains connected by this singlet rung are thus enlarged and compressed by $\delta M$ respectively.
The energy of this process comes mainly from the finite size corrections 
$\epsilon_b(M+\delta M)+\epsilon_b(M-\delta M)-2\epsilon_b(M)$ and is in the order of $M^{-3}$ (notice
that $\epsilon_b(M)-\epsilon_b(\infty)\sim M^{-1}$),
implying the larger the $M$, the more unstable the lattice.\par
 A ladder system may exhibit interesting behavior in an external magnetic field. 
In the TR phase, the system behaves as a two-component Luttinger liquid. The response of the system to the
magnetic field is rather usual, i.e., the zero field susceptibility shows a simple Pauli law.
 In the MS phase, a magnetic field will enlarge the period of the 
Wigner-like lattice and at a critical field $H_c$, the singlet rungs are no longer stable even in 
the ground state,
implying a phase transition between TR phase and MS phase. Two different situations may appear when a magnetic field
is applied on the RD phase. Roughly speaking, the magnetic field depresses the effect of $J$. Therefore,
with the increase of the field, the system flows either toward the TR phase or toward the MS phase. 
 If the system flows to the MS under a magnetic field, some 
flux phase may appear. Here the triplet rungs with $S_z=1$ serve as the fluxes in the singlet-rung
host. Since the ``repulsive interaction" $U$ occurs only between the nearest neighbor singlet rungs, 
the only possible flux phase has the structure $|0>_1\otimes |1>_2\otimes |0>_3\otimes \cdots \otimes
|0>_{N-1}\otimes |1>_{N}$, where $|0>_i$ and $|1>_j$ indicate the singlet rungs and the triplet
rungs, respectively and $N$ (even) represents the length of the ladder. With the increase of the external field,
the fluxes will clusterize and finally form a triplet-rung liquid or a completely polarized state.
\par
In conclusion, a generalized spin ladder is studied. It is found that in the extremely frustrated
case, a modulation structure which represents a Wigner-like lattice of the singlet rungs in the
triplet-rung host can exist in some parameter regions. Though only the $\beta=-1$ case is studied 
in detail, similar phenomena may exist for arbitrary $\beta$. The only difference is that the
TR domains in $\beta=-1$ case may be replaced by Haldane domains or VBS domains.\par
The author is  indebted to the hospitality
of Institut f\"{u}r Physik, Universit\"{a}t Augsburg, where this work was initiated.
 This work was partially supported by
AvH-Stiftung, NSFC, FCAS and NOYSFC.

\end{document}